\documentclass[aps,amsfonts,amsmath,prd,preprint,nofootinbib]{revtex4-1}
\newcommand{\beq}{\begin{equation}}
\newcommand{\eeq}{\end{equation}}

\usepackage{graphicx}
\usepackage{caption}
\usepackage{subcaption}

\def\beq{\begin{equation}}
\def\eeq{\end{equation}}
\def\beqa{\begin{eqnarray}}
\def\eeqa{\end{eqnarray}}

\def\){\right)}
\def\({\left(}
\def\]{\right]}
\def\[{\left[}

\begin{document}

\title{Stablizing oscillating universes against quantum decay}

\author{Audrey T. Mithani, Alexander Vilenkin}

\affiliation{Institute of Cosmology, Department of Physics and Astronomy,\\ 
Tufts University, Medford, MA 02155, USA}

\begin{abstract}

We investigate the effect of vacuum corrections, due to the trace anomaly and Casimir effect, on the stability of an oscillating universe with respect to decay by tunneling to the singularity.  We find that these corrections do not generally stabilize an oscillating universe.  However, stability may be achieved for some specially fine-tuned non-vacuum states.

\end{abstract}

\maketitle

\section{Introduction}

In recent years, there has been much interest in the possible existence of absolutely stable models describing a universe that can exist forever in a static or oscillating state.  Apart from the intrinsic interest of such models, they can play the role of an `eternal seed', providing a starting point for the emergent universe scenario (see, e.g., \cite{Mulryne,Sergio,Yu} and references therein).  One candidate model, dubbed the `simple harmonic universe' (SHU), has been suggested by Graham {\it et al} in Ref.~\cite{Graham1}.\footnote{SHU belongs to a more general class of models studied in Ref.~\cite{Dabrowski1}}  
The model is that of a closed spherical FRW universe,
%\beq
%ds^2 = dt^2 - a^2(t) d\Omega_3^2 ,
%\eeq
filled with a vacuum of negative energy density, $\Lambda < 0$, and with a material characterized by the equation of state $P=w\rho$ with $w=-2/3$ (this could be a network of domain walls).  
%Here, $a(t)$ is the scale factor and $d\Omega_3^2$ is the volume element on  a unit 3-sphere.
The corresponding Friedmann evolution equation is
\beq
\frac{{\dot a}^2 +k}{a^2} = \frac{1}{3} \rho(a) ,
\label{SHU}
\eeq
where $a(t)$ is the scale factor, $k=+1$ is the curvature parameter, the energy density $\rho(a)$ is given by
\label{Friedmann}
\beq
\rho(a) = \Lambda +\frac{\sigma}{a} ,
\label{rhoa}
\eeq
and we use Planck units in which $8\pi G = 1$.  
%For a domain wall network, $\sigma$ has the meaning of mass per unit area of the wall.

With a suitable choice of parameters, the model is perturbatively stable \cite{Graham1,MV2,Graham2}, but it was shown in \cite{MV1} (following the earlier work of \cite{Dabrowski}) that it is unstable with respect to quantum tunneling to a singular state of zero volume $(a=0)$.   In classical theory, the model can be regarded as a constrained dynamical system with a Hamiltonian
\beq
{\cal H}=-\frac{1}{12\Omega a} \left( p_a^2 +U(a)\right) ,
\label{H}
\eeq
where 
\beq
p_a = -6\Omega a{\dot a}
\label{pa}
\eeq
is the momentum conjugate to $a$ and
\beq
U(a) = (6\Omega)^2 a^2 \left( k -\frac{1}{3} a^2 \rho(a) \right).
\label{Ua}
\eeq
Here, $\Omega$ is the comoving volume of the universe, that is, the volume at $a = 1$.  For a spherical universe, 
\beq
\Omega(k=1)=2\pi^2.  
\eeq
A universe with $k\neq 1$ can also have a finite volume in the case of nontrivial topology.  We have included this case for later discussion.  For all values of $k$, the Hamiltonian constraint ${\cal H}=0$ yields the evolution equation (\ref{SHU}).  

The effective potential $U(a)$ for the model (\ref{rhoa}) with $k=1$ is plotted in Fig.~1(a).  The universe oscillates between the classical turning points
\beq
a_{\pm} = \frac{\sigma}{2|\Lambda|} \( 1 \pm \sqrt{1-\frac{12|\Lambda|}{\sigma^2}} \),
\label{apm}
\eeq
but it can also tunnel through the barrier from $a_-$ to the singularity at $a=0$.  It can be easily verified that the WKB tunneling action 
\beq
S=\int_0^{a_-} |p_a| da = \int_0^{a_-} \sqrt{U(a)} 
\eeq
is finite.  The quantity
\beq
{\cal P} ~ e^{-2S}
\eeq
can be interpreted as the probability of collapse through quantum tunneling as the universe bounces at radius $a=a_-$.
\begin{figure}[t]
\centering
\begin{subfigure}{.49\textwidth}
\includegraphics[width=\linewidth]{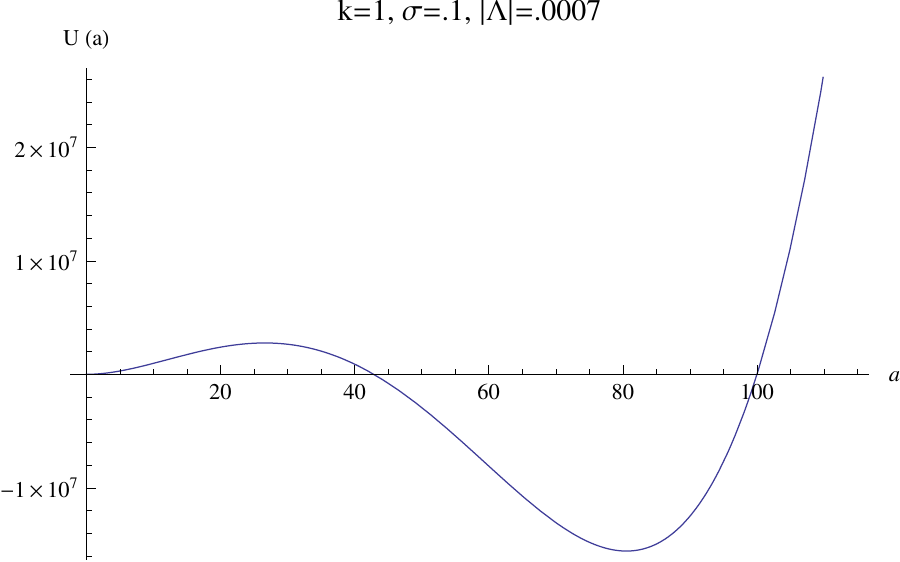}
\caption{}
\label{potential}
\end{subfigure}%
         %add desired spacing between images, e. g. ~, \quad, \qquad, \hfill etc.
          %(or a blank line to force the subfigure onto a new line)
\begin{subfigure}{.49\textwidth}
\includegraphics[width=\linewidth]{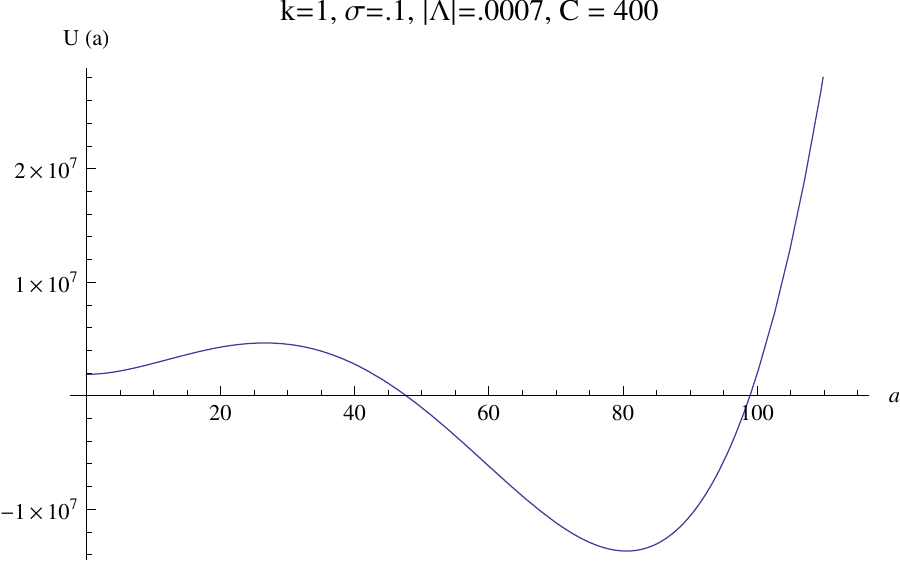}
\caption{}
\label{potentialC}
\end{subfigure}
\caption{The effective potential $U(a)$ of the simple harmonic universe (a), and the simple harmonic universe with Casimir energy (b).}
\end{figure}

Most recently, Graham {\it et al} suggested an interesting possibility that the SHU model can be stabilized by Casimir energy due to the zero-point fluctuations of quantum fields \cite{Graham2}.  They assume that the Casimir energy density is given by
\beq
\rho_C(a) = -\frac{C}{a^4}
\label{rhoC}
\eeq
with $C>0$.  If $\rho_C(a)$ is added to the energy density in Eq.~(\ref{rhoa}), the potential $U(a)$ in (\ref{Ua}) takes the form shown in Fig.~1(b).  The minimum of the potential at $a=0$ is lifted to a positive value, and Graham {\it et al} argue that this should prevent the tunneling from taking place.

In the present paper we shall examine, in some more detail, the effect of Casimir and other quantum corrections to the energy-momentum tensor on the tunneling of the universe.  A Casimir energy density of the form (\ref{rhoC}) was derived for massless quantum fields in a static Einstein universe.  In an oscillating universe like the SHU, there are additional contributions, depending on time derivatives of the scale factor.  Even in the static limit, when $a_+ = a_-$, the tunneling is described by an instanton solution $a(\tau)$, which depends on the Euclidean time $\tau$, so the additional terms in $T_{\mu\nu}$ must be included.
We shall see that these terms can have a significant effect on the tunneling.

Even when keeping only the Casimir energy is a good approximation, we shall argue that it does not generally prevent the universe from quantum decay.  It may be possible, however, to stabilize the universe in some finely-tuned non-vacuum states.

%The paper is organized as follows.  [........]

\section{Energy-momentum tensor}

Calculation of the expectation value of the energy momentum tensor $\langle T^q_{\mu\nu}\rangle$ of quantum fields in a curved spacetime is a rather challenging task and can be done in a closed form only in a few simple cases (see \cite{Birrell,Grib} for a review).  The case most relevant to our considerations, which we shall adopt here, is that of free, massless, conformally coupled fields in a FRW universe.  Then $\langle T^q_{\mu\nu}\rangle$ can be represented as
\beq
\langle T^{q}_{\mu\nu}\rangle = \alpha ~ ^{(1)}H_{\mu\nu} + \beta ~ ^{(3)}H_{\mu\nu} + T^{(C)}_{\mu\nu} + T^{(s)}_{\mu\nu}  ,
\label{Tmunu} 
\eeq
where $R_{\mu\nu}$ is the Ricci tensor and we have used the standard notation \cite{Birrell}
\beq
^{(1)}H_{\mu\nu} = 2R_{;\mu;\nu} -2g_{\mu\nu}R_{;\sigma}^{:\sigma} +2RR_{\mu\nu}-\frac{1}{2}g_{\mu\nu}R^2 
\label{H1} 
\eeq
and
\beq
^{(3)}H_{\mu\nu} = R_\mu^\sigma R_{\nu\sigma} - \frac{2}{3}RR_{\mu\nu} -\frac{1}{2}g_{\mu\nu}R_{\sigma\tau}R^{\sigma\tau} +\frac{1}{4}g_{\mu\nu}R^2 .
\label{H3}
\eeq
The coefficient $\beta$ in (\ref{Tmunu}) is determined by the trace anomaly; it is given by
\beq
\beta = \frac{1}{1440\pi^2}(N_0 + 11 N_{1/2} +31 N_1),
\label{H0}
\eeq
where $N_0$, $N_{1/2}$ and $N_1$ are the numbers of quantum fields of spin 0, 1/2 and 1, respectively.  (Note that $N_{1/2}$ is the number of chiral spinors; a Dirac spinor is counted as two chiral spinors.)
The tensor $^{(3)}H_{\mu\nu}$ is not identically conserved, but it is conserved in conformally flat spacetimes (and thus in FRW spacetimes).  This tensor cannot be obtained by varying a local action.  On the other hand, the tensor $^{(1)}H_{\mu\nu}$ can be obtained by varying an $R^2$ term in the action and is identically conserved.  The coefficient $\alpha$ is affected by the $R^2$ counterterm that has to be added in order to cancel infinities in $T_{\mu\nu}$.\footnote{In the Starobinsky model of inflation \cite{Starobinsky79}, the coefficient $\beta$ determines the expansion rate, and the $\alpha$ term gives rise to an effective scalar degree of freedom of tachyonic mass $i(6\alpha)^{-1/2}$.  This term is responsible for the exit from inflation.  In order to have long inflation, Starobinsky assumed that $\alpha\gg \beta$.}   By a suitable choice of the counterterm, this coefficient can be tuned to zero.  We shall adopt this choice here, in order to simplify the discussion.

%but the trace anomaly contribution (proportional to $^{(3)}H_{\mu\nu}$) cannot be avoided.  

$T^{(C)}_{\mu\nu}$ is an additional Casimir contribution, which arises if the space has nontrivial topological identifications.  One example is a spatially flat universe with a toroidal topology; then
\beq
T^{(C)}_{\mu\nu}  = -\frac{C}{a^4} {\rm diag} (1, 1/3, 1/3, 1/3).
\label{Casimir}
\eeq
This tensor is traceless and covariantly conserved: $T^{(C)\nu}_\nu = 0$, $T^{(C)\nu}_{\mu ;\nu} =0$.
The coefficient $C$ for a real scalar field is $C_0 = 0.8375$.  In general, for non-interacting conformal fields,
\beq
C = (N_b - N_f)C_0 ,
\label{C}
\eeq
where $N_b = N_0 + 2N_1$ and $N_f = 2N_{1/2}$ are the numbers of bosonic and fermionic spin degrees of freedom.\footnote{Note that $T^{(C)}_{\mu\nu}$ vanishes in supersymmetric models, where  
$N_b=N_f$.}  A flat toroidal universe will be discussed in detail in Sec.~III.  Note that for a 
spherical $(k=+1)$ universe, there are no additional Casimir terms: the Casimir contribution is already included in $^{(3)}H_{\mu\nu}$.  The reason is that a $k=1$ universe is conformally related to a static Einstein universe, which has a vanishing trace anomaly.

The last term in (\ref{Tmunu}) depends on the choice of quantum state; it can be thought of as `radiation', representing particle excitations of the fields.  For conformal fields, there is a natural choice of vacuum state -- the conformal vacuum, whose mode functions
are obtained by a conformal transformation from those of a static universe with $a={\rm const}$.  In this state $T^{(s)}_{\mu\nu}=0$.  (Note that for conformal fields the time dependence of the scale factor does not give rise to particle production.)  For now, we shall assume that all fields are in their conformal vacuum states; a more general case will be considered in Sec.~IV.

To investigate the effect of quantum corrections on the SHU, we shall use the semiclassical Einstein equations,
\beq
R_{\mu\nu} -\frac{1}{2}g_{\mu\nu}R =  -T^{SHU}_{\mu\nu}-\langle T^q_{\mu\nu}\rangle ,
\label{semiclassical}
\eeq
where $T^{SHU}_{\mu\nu}$ is the classical energy-momentum tensor of the SHU.  The corresponding 
Friedmann equation is 
\beq
\frac{\dot a^2 +k}{a^2} = \beta \( \frac{\dot a^2 +k}{a^2}\)^2 
%- \frac{1}{M^2}\( 2\frac{\dot a \dddot a}{a^2} - \frac{\ddot a^2}{a^2} + 2\frac{\ddot a \dot a^2}{a^3} -3\frac{\dot a^4}{a^4} \)
+ \frac{1}{3}\left(\Lambda +\frac{\sigma}{a}\right) , 
\label{Friedmann}
\eeq
where we have set $\alpha = 0$ and $T^{(C)}_{\mu\nu}=0$
(assuming spherical topology).

As we already mentioned, the trace anomaly term in this case includes the effect of Casimir energy (proportional to $a^{-4}$ in the Friedmann equation).  We see, however, that this energy is positive, so its effect is to lower the potential, creating another classically allowed region near $a=0$.

We can find the classical turning points by setting $\dot a = 0$ in the Friedmann equation (\ref{Friedmann}), and finding solutions for
\beq
1 - \frac{1}{3}\( \sigma a - \Lambda a^2Ê\) =Ê \beta a^{-2}.
\eeq
With a suitable choice of the parameters, the classical turning points (\ref{apm}) remain approximately unchanged.  (The condition for that is $\beta |\Lambda|^3/ \sigma^4 \ll 1$.)  Then it is easy to see that there must be an additional turning point at small $a$.  With $\beta\sigma^2  \ll 1$, this turning point is at
\beq
a_*\approx \beta^{1/2}.
\eeq
If the number of quantum fields appearing in Eq.~(\ref{H0}) is sufficiently large, $N \gg 10^3$, so that $\beta \gg 1$, then $a_*$ is large in Planck units.  There is then a classically allowed region between $a_*$ and $0$, and clearly the oscillating universe can tunnel to this region.

Graham {\it et al} have pointed out that a negative Casimir energy for a $k=+1$ SHU may be possible to arrange by considering non-conformal fields.  The case of a massless non-conformally coupled scalar field was studied in Ref.~\cite{Herdeiro}, with the conclusion that the Casimir energy density is given by Eq.~(\ref{rhoC}), where the sign of the coefficient $C$ depends on the coupling $\xi$ of the field to the curvature.  It is negative for conformal coupling, $\xi=1/6$, but is positive for the minimal coupling, $\xi=0$.  Unfortunately the calculation of $\langle T_{\mu\nu}\rangle$ in Ref.~\cite{Herdeiro} was performed only for a static Einstein universe, which is not sufficient for the analysis of tunneling.  In the next section, we shall therefore introduce a version of the SHU model which allows a negative Casimir energy for conformally coupled fields, so that Eq.~(\ref{Tmunu}) can still be used.

\section{A flat oscillating model} 

We consider a spatially flat $(k=0)$ universe,
\beq
ds^2 = dt^2 - a^2(t) d{\bf x}^2 ,
\eeq
which is compactified on a torus: the coordinates $x^i$ $(i=1,2,3)$ take values in the range $0\leq x^i \leq 1$ and the surfaces $x^i =0$ and $x^i =1$ are identified.  In this case the expectation value of $T_{\mu\nu}$ has an additional Casimir contribution  $T^{(C)}_{\mu\nu}$, given by Eqs.~(\ref{Casimir}),(\ref{C}).  With $N_b > N_f$, we have $C>0$ and the Casimir energy is negative.
The $k=0$ Friedmann equation with the trace anomaly and Casimir contributions is
\beq
\frac{\dot a^2}{a^2} = \beta \( \frac{\dot a^2}{a^2}\)^2+ \frac{1}{3} \( \Lambda + \frac{\sigma}{a} - \frac{C}{a^4}\).
\label{FriedmannFlat}
\eeq
The turning points are the solutions to the Friedmann equation when $\dot a=0$:
\beq
\Lambda a^4 + \sigma a^3 - C =0.
\eeq
Though we cannot find analytic solutions, there are two classical turning points for $C < \frac{27 \sigma^4}{256  |\Lambda|^3}$.  For $\sigma^4 /|\Lambda|^3 \gg C$, the turning points are approximately given by
\beq
a_+\approx \sigma/|\Lambda| , ~~~ a_-\approx (C/\sigma)^{1/3}.
\label{a+a-}
\eeq

Following Graham {\it et al} \cite{Graham2}, we shall first consider the regime where the trace anomaly term can be neglected.  (The effect of this term will be discussed in the next section.)  
The Friedmann equation then corresponds to the Hamiltonian constraint ${\cal H}=0$ with ${\cal H}$ from Eq.~(\ref{H}) and $\Omega(k=0)=1$, so the potential is given by
\beq
U(a) = 12 \(-\sigma a^3- \Lambda a^4 + C\) .
\eeq
The Casimir contribution to the potential is a positive constant, so $U(a)$ remains positive all the way from $a_-$ to $a=0$ and has the general form illustrated in Fig.~1(b).  Graham {\it et al} argued that in this case the universe should be stable with respect to quantum decay.  

To examine this claim, we consider the wave function of the universe, which can be found by solving the Wheeler-DeWitt (WDW) equation\footnote{Here we assume the simplest ordering of the operators $a$ and $p_a$ in the Hamiltonian (\ref{H}).}
\beq
\(\frac{d^2}{da^2} - U(a)\)\psi(a)=0.
\label{WDW}
\eeq
The probabilistic interpretation of the wave function $\psi$ is still a matter of some debate.  Some authors \cite{interpretation,Halliwell} suggest that the probability density should be expressed in terms of the current,
\beq
J = i(\psi^*\nabla\psi - \psi\nabla\psi^*) ,
\eeq
where the gradient $\nabla$ is with respect to the superspace variables.  This approach has a number of attractive properties, but it fails in the case of a stationary state, like the one we discuss here, when the wave function is real and $J=0$.  An alternative approach \cite{HP1} is the `naive Schrodinger measure',
\beq
{\cal P} \propto |\psi|^2 .
\label{psi2}
\eeq
It has been argued in \cite{HP1} that the two approaches agree under certain assumptions.  We will not attempt to resolve this issue here and will simply adopt the naive definition (\ref{psi2}).

In our model, the probability for the universe to have infinite size should vanish; hence we have to require that
\beq
\psi(a\to\infty)=0 .
\label{bc}
\eeq
In order to exclude the singular state at $a=0$, one would also want to require $\psi(0)=0$ \cite{DeWitt}.  However, the general solution of Eq.~(\ref{WDW}) depends on only two arbitrary constants.  One of them is used to fix the overall normalization of $\psi$ and the other to enforce the boundary condition (\ref{bc}).  Thus there is no freedom left to enforce a boundary condition at $a=0$ \cite{MV1}.    

The WKB solutions for $\psi$ in the classically forbidden range $0<a<a_-$ have the form
\beq
\psi_\pm(a) \propto \( U(a)\)^{-1/4} e^{\pm W(a)} ,
\eeq
where
\beq
W(a) = \int_a^{a_-} \sqrt{U(a)} da .
\eeq
The wave function will generally include a superposition of $\psi_+$ and $\psi_-$.  Since $\psi_+$ grows exponentially with decreasing $a$, we expect $\psi$ to be large at $a=0$.  This behavior is illustrated in a numerical solution shown in Fig.~2.

\begin{figure}[t]
\centering
\includegraphics[width=10cm]{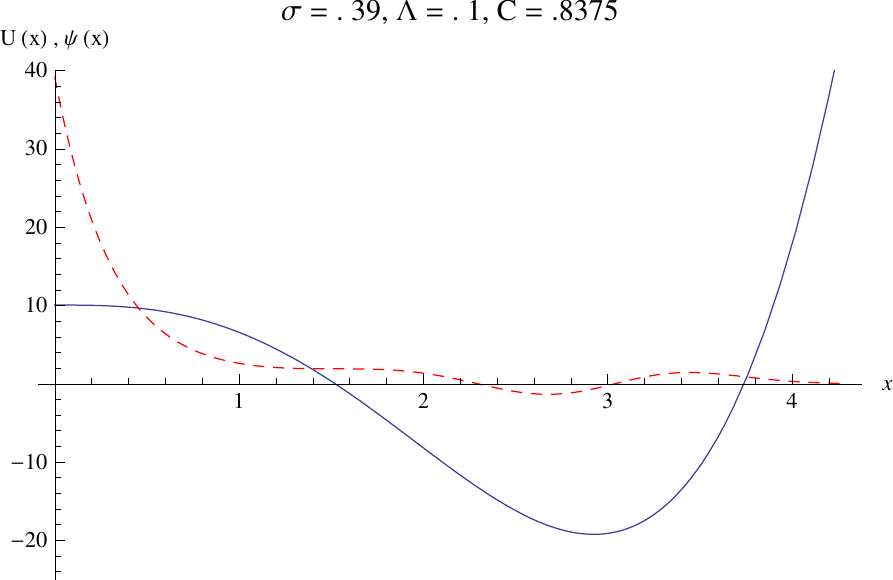}
\caption{Numerical solution (dashed red line) to the WDW equation with $\sigma = .39$, $\Lambda = -.1$ and $C = .8375$.  The potential is plotted with a solid blue line.}  
\end{figure}

The fact that $\psi(0)$ is large indicates that the UV physics at $a\lesssim 1$ cannot be ignored.  Without a UV-complete theory of quantum gravity, we cannot tell what the effect of this physics will be, but we shall try to consider some possible alternatives.  If reaching $a\lesssim 1$ means disappearance of semiclassical spacetime, then a stationary state with $\psi(0)\neq 0$ simply cannot exist.  To illustrate this point, let us consider a hypothetical world where the electron wave function in a hydrogen atom is such that the probability density for the electron to be at the location of the proton is nonzero. Suppose further that
the proton, electron, neutron and neutrino masses satisfy $m_p + m_e >  m_n +m_\nu$, so that proton and electron can always scatter into neutron and neutrino, 
\beq
e^- p^+ \to n\nu .
\label{epnnu}
\eeq  
Under these assumptions, it is clear that the atom would have a nonzero decay rate.  We can imagine that the interaction between electrons and protons in this world is such that the electron `orbit' is separated from the center of the atom by a potential barrier.   There may or may not be a small classically allowed range near the center.  Our conclusion applies in either case: a stationary state of the atom is not possible if the probability density for the electron at the center is nonzero.
In this example, the reaction (\ref{epnnu}) represents the UV physics, which is not included in the Schrodinger equation for the hydrogen atom, just like the disintegration of the classical spacetime at $a=0$ is not reflected in the WDW equation.      

An alternative scenario is that the UV-complete theory will resolve the singularity at $a=0$, replacing it with a non-singular Planck-size nugget.  The universe may then tunnel back and forth between the classical oscillating regime and the nugget, resulting in a stationary quantum state.  The WDW equation should be accurate in the semiclassical regime at $a\gg 1$; hence our conclusion regarding the growth of the wave function towards small values of $a$ should still apply.  This indicates that the most probable states of the universe will be in the Planck regime at $a\sim 1$.

\section{Stable oscillating universes}

Even though the boundary condition $\psi(0)=0$ cannot be generally enforced, it may be satisfied for some special values of the parameters $\Lambda$, $\sigma$ and $C$.  These parameters are assumed to be fixed, but the effective value of $C$ can be changed if we allow more general states of the quantum fields contributing to the Casimir energy.  We assumed so far that these fields are all in conformal vacuum states, so that $T^{(s)}_{\mu\nu}=0$ in Eq.~(\ref{Tmunu}).  Suppose now that some particle excitations on top of the vacuum are also present.  For massless particles, the expectation value of $T_{\mu\nu}$ will have the form of the radiation energy-momentum tensor,
\beq
T^{(s)}_{ \mu\nu}  = \frac{C_r}{a^4} {\rm diag} (1, 1/3, 1/3, 1/3)
\label{radiation}
\eeq
with $C_r > 0$.  This has the same form as the Casimir term (\ref{Casimir}).  Hence, including particle excitations has the effect of replacing the parameter $C$ with an effective value
\beq
C_{eff} = C - C_r .
\eeq

Increasing $C_r$ (decreasing $C_{eff}$) widens the potential well and increases the number of nodes of the wave function in the classically allowed region, causing $\psi(0)$ to oscillate between positive and negative values and pass through zero at special values of $C_{eff}$.  We can then tune the particle content in such a way that $C_{eff}$ takes one of the values that yield $\psi(0)=0$.  For a given set of parameters $\Lambda$, $\sigma$, and $C$, there is a finite number of such nonsingular solutions, given by the difference in the number of nodes at $C_r = C$ and at $C_r = 0$.

As an illustration, we show in Fig.~3 the non-singular solution for the parameter values $\sigma = .39$, $\Lambda = -.1$ and $C = .8375$ (the same as in Fig.~2).  There are only two such solutions in this case, corresponding to the values $C_r = .6289, .09506$.

\begin{figure}[t]
\centering
\begin{subfigure}{.49\textwidth}
\includegraphics[width=\linewidth]{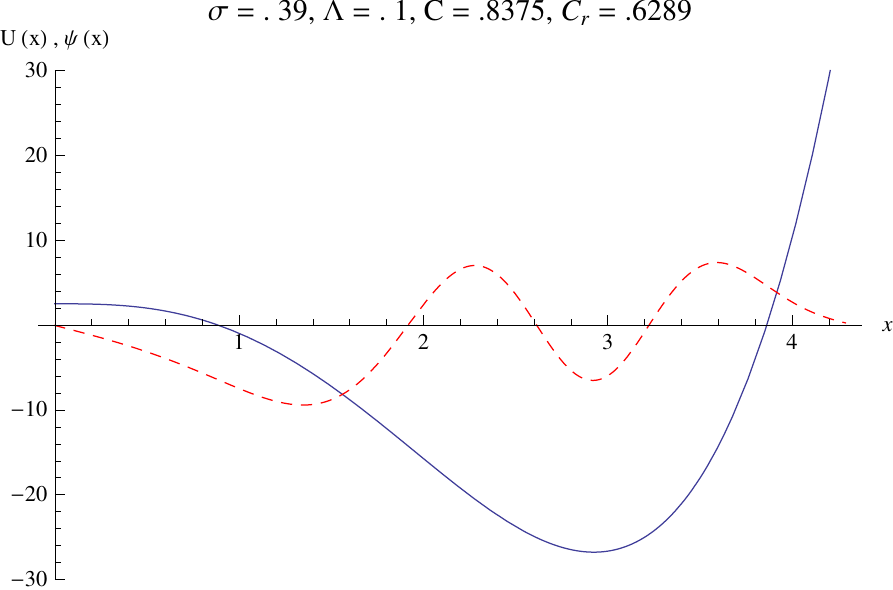}
\caption{$C_r = .6289$}
\end{subfigure}
\begin{subfigure}{.49\textwidth}
\includegraphics[width=\linewidth]{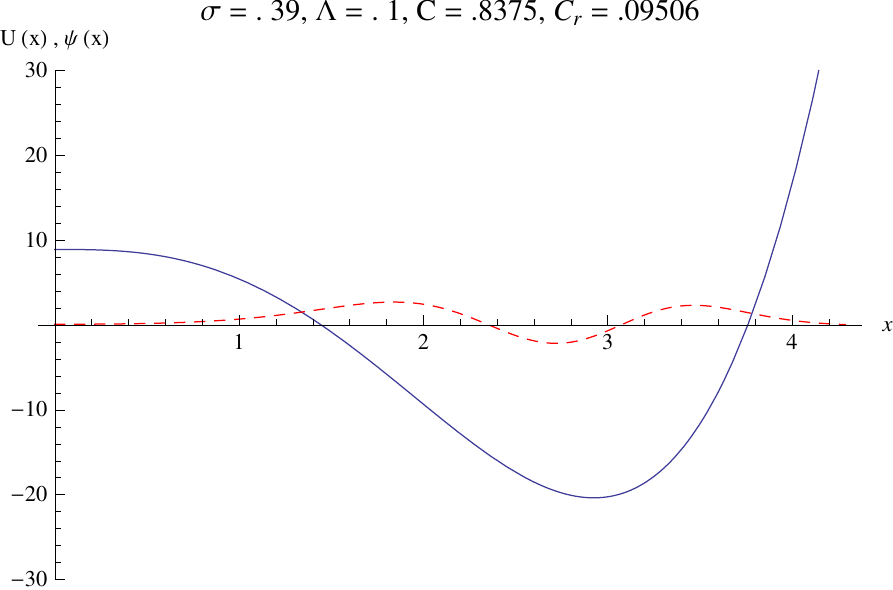}
\caption{$C_r =  .09506$}
\end{subfigure}
\caption{Numerical solutions of the WDW equation (dashed red line) with parameters $\sigma = .39$, $\Lambda = -.1$ and $C = .8375$, and two values of $C_r$ such that $\psi(0) =0$.  The potential is plotted with a solid blue line.}  
\label{stable}
\end{figure}

\section{The effect of the trace anomaly}

Let us now consider the effect of the trace anomaly term (proportional to $\beta$) in Eq.~(\ref{FriedmannFlat}).  The effective Lagrangian corresponding to this equation is \cite{Hartle,Barvinsky}
\beq
{\cal L} = -3a{\dot a}^2 +\beta \frac{{\dot a}^4}{ a} - a^3 \rho(a) ,
\eeq
where
\beq
\rho(a)=\Lambda+\frac{\sigma}{a} -\frac{C}{a^4} .
\eeq
Wheeler-DeWitt quantization of this model is problematic, because the momentum $p_a = \partial {\cal L} /\partial {\dot a}$ depends nontrivially on ${\dot a}$, and as a result the Hamiltonian operator involves fractional powers of differential operators.  We shall therefore take an alternative route and analyze tunneling in terms of solutions to Euclideanized Friedmann equation.

With the replacement $t\to -i\tau$, the Friedmann equation (\ref{FriedmannFlat}) takes the form
\beq
\frac{{a'}^2}{a^2}+\beta \frac{{a'}^4}{a^4} +\frac{1}{3} \rho(a) = 0 ,
\eeq
where primes stand for derivatives with respect to the Euclidean time $\tau$.  Solving this for $a'$, we have
\beq
{a'}^2 = \frac{a^2}{2\beta} \( -1 \pm \sqrt{1 - \frac{4\beta}{3} \rho(a)} \) .
\label{a'}
\eeq
In the classically forbidden range $\rho(a) < 0$, so we have to choose the positive sign of the square root.
Then it is easy to check that for $|\rho(a)| \ll \beta^{-1}$ the trace anomaly term is unimportant and Eq.~(\ref{a'}) reduces to the usual (Euclidean) Friedmann equation,
\beq
{a'}^2 = -\frac{1}{3} \rho(a) .
\eeq
At small $a$, $\rho(a)$ is dominated by the Casimir term, $\rho_C(a) = -C/a^4$, and the condition $|\rho(a)| \ll \beta^{-1}$ gives
\beq
a \gg ( \beta C)^{1/4} \equiv a_q .
\eeq

For generic numbers of quantum fields, $N_0\sim N_{1/2} \sim N_1 \sim N$, $a_q \sim 0.2 N^{1/2}$, so we can have $a_q \gg 1$ if $N$ is sufficiently large.  We note also that with sufficiently small $\sigma$ Eq.~(\ref{a+a-}) gives $a_- \gg a_q$.  Thus, with a suitable choice of parameters the scale factor ranges $1\ll a \lesssim a_q$ and $a_q \ll a \lesssim a_-$ should both allow semiclassical treatment.

In the range $a\ll a_q$, where the trace anomaly term is important, Eq.~(\ref{a'}) takes the form
\beq
{a'}^2 \approx (C/3\beta)^{1/2} ,
\eeq
with the solution
\beq
a(\tau) \approx (C /3\beta)^{1/4} \tau .
\label{atau}
\eeq
The Euclidean action evaluated in this range is
\beq
S_E(\tau) =  \int_\tau^{\tau_-} d\tau \( 3a{a'}^2 +\frac{\beta{{a'}^4}}{a} - a^3 \rho(a)\) ,
\label{SE}
\eeq
where $\tau_- \sim C^{1/12} \beta^{1/4}\sigma^{-1/3}$ corresponds to $a=a_-$.  We see immediately that with $a(\tau)$ from (\ref{atau}) the integral in (\ref{SE}) diverges at $\tau\to 0$.  Expressing $\tau$ in terms of $a$, we have
\beq
S_E(a\to 0) \approx  Q \ln \frac{a_-}{a} ,
\eeq
where
\beq
Q = (3\beta C^3)^{1/4}  \sim N .
\eeq
This suggests that the wave function at small $a$ grows as a large negative power of $a$,
\beq
\psi(a) \propto e^{S_E(a)} \propto a^{-Q} .
\eeq
The growth may or may not be terminated by quantum gravity effects at $a\sim 1$.  In either case, normalizable wave functions with $\psi(a\to 0)=0$ may exist for some special states of the quantum fields, as discussed in Sec.~IV.

\section{Summary and discussion}

We used a simple minisuperspace model to analyze the effect of Casimir and trace anomaly corrections on the quantum decay of classically stable oscillating universe models.  We found that these corrections can significantly modify the wave function of the universe $\psi(a)$ at small values of the scale factor $a$ and may even cause it to diverge at $a\to 0$.  However, the vacuum corrections do not generally stabilize an oscillating universe.  The reason is very simple: the wave function of the universe must satisfy the boundary condition $\psi(a\to\infty)=0$.  This leaves no freedom to impose a boundary condition at $a=0$, and as a result the wave function generally grows towards small $a$.  In this regard the situation is the same as for the simple harmonic universe model without vacuum corrections \cite{MV1}.

We found also that the wave function can be tuned to zero at $a=0$ for non-vacuum states of the quantum fields, corresponding to certain fine-tuned amounts of `radiation' in the universe.  Such states may correspond to absolutely stable, stationary quantum states of the universe.

This possibility, however, should be regarded with caution.  Our treatment here has been based on the semiclassical gravity approximation, Eq.~(\ref{semiclassical}), in which the quantity $C_r$ characterizing the amount of radiation in Eq.~(\ref{radiation}) is a continuous parameter.  This allowed us to tune this parameter to enforce $\psi(a=0)=0$.  It is not clear that such tunable parameters will exist in the full theory of quantum gravity.   One might expect that, on the contrary, $C_r$ could be quantized.  For example, if we simply add
quantum conformal free fields to a compact minisuperspace FRW model, the WDW equation will separate and 
the fields will contribute a radiation term with a discrete spectrum of $C_r$, which is not likely to overlap with the set of values required for $\psi(0)=0$.  This treatment, however, would not account for the Casimir and trace anomaly contributions and for the possibility of having the quantum fields in a superposition of occupation number eigenstates.  A definitive resolution of these issues will require a better understanding of the quantum theory of gravity. 

We finally note an interesting result obtained recently by Damour and Spindel in Ref.~\cite{Damour}.  They developed a quantum theory of a supersymmetric Bianchi IX model and found that this model admits quantum states satisfying $\psi(a=0)=0$ without any additional fine-tuning.

\section*{Acknowledgements}

We are grateful to Alan Guth, Ben Freivogel and Gary Gibbons for useful discussions.  This work was supported in part by the National Science Foundation (grant PHY-1213888).

\end{document}